\documentclass[doublecol]{epl2}

\title{Dynamic instability of cooperation due to diverse activity patterns in evolutionary social dilemmas}
\shorttitle{Dynamic instability of cooperation}
\author{Cheng-Yi Xia\inst{1,2} \and Sandro Meloni\inst{2,3} \and Matja{\v z} Perc\inst{4,5} \and Yamir Moreno\inst{2,3,6}}
\shortauthor{Xia et al.}

\institute{\inst{1}Key Laboratory of Computer Vision and System and Tianjin Key Laboratory of Intelligence Computing and Novel Software Technology, Tianjin University of Technology, Tianjin 300191, People's Republic of China\\
\inst{2}Institute for Biocomputation and Physics of Complex Systems, University of Zaragoza, E-50018 Zaragoza, Spain\\
\inst{3}Department of Theoretical Physics, University of Zaragoza, E-50009, Zaragoza, Spain\\
\inst{4}Faculty of Natural Sciences and Mathematics, University of Maribor, Koro{\v s}ka  cesta 160, SI-2000 Maribor, Slovenia\\
\inst{5}Department of Physics, Faculty of Science, King Abdulaziz University, Jeddah, Saudi Arabia\\
\inst{6}Complex Networks and Systems Lagrange Lab, Institute for Scientific Interchange, Turin 10126, Italy}


\abstract{Individuals might abstain from participating in an instance of an evolutionary game for various reasons, ranging from lack of interest to risk aversion. In order to understand the consequences of such diverse activity patterns on the evolution of cooperation, we study a weak prisoner's dilemma where each player's participation is probabilistic rather than certain. Players that do not participate get a null payoff and are unable to replicate. We show that inactivity introduces cascading failures of cooperation, which are particularly severe on scale-free networks with frequently inactive hubs. The drops in the fraction of cooperators are sudden, while the spatiotemporal reorganization of compact cooperative clusters, and thus the recovery, takes time. Nevertheless, if the activity of players is directly proportional to their degree, or if the interaction network is not strongly heterogeneous, the overall evolution of cooperation is not impaired. This is because inactivity negatively affects the potency of low-degree defectors, who are hence unable to utilize on their inherent evolutionary advantage. Between cascading failures, the fraction of cooperators is therefore higher than usual, which lastly balances out the asymmetric dynamic instabilities that emerge due to intermittent blackouts of cooperative hubs.}

\begin{document}

\maketitle
Evolutionary game theory \cite{maynard_82, nowak_06} is the theoretical framework of choice for studying the evolution of cooperation in social dilemmas \cite{axelrod_84}. In evolutionary games strategies with a high fitness proliferate, while strategies who fail to reproduce are destined to disappear through natural selection. The problem in the realm of social dilemmas is that the strategy that ensures the highest individual fitness is defection, while the society as a whole is best off if everybody would cooperate. The expectation therefore is that cooperative behavior should not have evolved under such circumstances. Yet cooperation in nature abounds \cite{wilson_71, nowak_11, rand_tcs13}, even under circumstances that constitute a social dilemma, and it is difficult to reconcile this with the fundamental assumptions of natural selection that are described in Darwin's \textit{The Origin of Species}. Indeed, widespread altruistic cooperation is an important challenge to Darwin's theory of evolution, and knowing how it evolved is fundamental for the understanding of the main evolutionary transitions that led from single-cell organisms to complex animal and human societies \cite{maynard_95}.

Due to the universal appeal of the subject, ample research efforts have been invested towards reconciling theory and observations, and several mechanisms have been identified that promote cooperation in evolutionary social dilemmas. While Hamilton's kin selection theory \cite{hamilton_wd_jtb64a} and various forms of reciprocity and group selection are probably the most famous examples \cite{nowak_s06}, many refinements to these fundamental mechanisms have been proposed and explored. Evolutionary games in structured populations, in particular, have recently attracted considerable attention \cite{szabo_pr07, roca_plr09, perc_bs10, santos_jtb12, perc_jrsi13}, as it becomes clear that methods of statistical physics \cite{binder_88, castellano_rmp09} can be employed successfully to study the evolution of cooperation on networks.

It has been discovered that the emergence of cooperation and the phase transitions leading to other favorable evolutionary outcomes depend sensitively on the structure of the interaction network and the type of interactions, as well as on the number and type of competing strategies \cite{abramson_pre01, szabo_prl02, zimmermann_pre04, santos_prl05, ohtsuki_prl07, gomez-gardenes_prl07, pena_pre09, szolnoki_prl12, szolnoki_prx13, tanimoto_pre13}. Theoretical studies that are unique to physicists have led to significant advances in our understanding of the evolution of cooperation, for example by revealing the importance of heterogeneity of interaction networks \cite{santos_prl05} or the dynamical organization of strategies \cite{gomez-gardenes_prl07}, to name just two examples. Keeping pace with theoretical advances are also human experiments \cite{grujic:2010, rand_2011, wang_2012, gracia_2012b, bednarik_14, rand_14}, which are becoming more and more common.

An important assumption behind existing research, however, has been that players always participate in each round of the game \cite{perra_scirep2012}. In reality, however, players will often abstain from competition, for example due to lack of interest or to mitigate risk. In this letter, we therefore consider evolutionary social dilemmas where the participation of each player at a particular round of the game is probabilistic rather than certain. We note that our consideration of inactivity is different from the consideration of loners \cite{szabo_prl02}. Loners refuse to participate and rather rely on some small but fixed income, yet they are always active and can replicate. In our case, an inactive player always gets a null payoff and is unable to replicate. Moreover, we consider inactivity as a dynamically changing state of players rather than a strategy. As we will show, inactivity introduces cascading failures of cooperation, which only under special circumstances are compensated by the inactivity-induced inability of defectors to utilize on their inherent evolutionary advantage over cooperators.

Formally, we study pairwise evolutionary games on Erd\H{o}s-R\'enyi (ER) random networks and on scale-free (SF) networks generated according to the uncorrelated configuration model \cite{catanzaro_pre05}, each with an average degree $k=4$ and size $N=10^{4}$. At this point, we note that the consideration of different activity patterns does not affect the expected outcome of social dilemmas in well-mixed populations. During each pairwise interaction, mutual cooperation yields the reward $R$, mutual defection leads to punishment $P$, and the mixed choice gives the cooperator the sucker's payoff $S$ and the defector the temptation $T$. Within this setup we have the commonly employed weak prisoner's dilemma game \cite{nowak_n92b} if $T>R>P=S$. Without loss of generality we use $R=1$, $P=S=0$ and $T = b \ge 1$ as a representative setup for a social dilemma. Cooperators ($C$) and defectors ($D$) are initially distributed uniformly at random, so that they have equal chances for success during evolution.

Simulations are performed in agreement with a synchronous updating protocol, such that each player $i$ plays the game with all its $k_i$ neighbors and thereby collects the payoff $\Pi_i$. Once all the players collected their payoffs an evolutionary step take place. For the evolution of strategies, we employ the replicator dynamics, such that if $\Pi_j > \Pi_i$, player $j$ will replicate its strategy $s_j$ to the site of player $i$ with the probability
\begin{equation}
P_{s_j \rightarrow s_i}= \frac{\Pi_j-\Pi_i}{\mbox{max}(k_j,k_i)b}.
\end{equation}
We perform simulations until the average fraction of cooperators in the population $\langle C \rangle$ becomes time independent, that is, until a stationary state has been reached.

To introduce different activity patterns, we assign to every player an activation probability $a_i \in [0,1]$ according to which player $i$ participates in a particular round of the game. Accordingly, with probability $1-a_i$ the player $i$ remains inactive in each round of the game. If a player $i$ is active, it will play the game with all its $k_i$ neighbors, independently from whether they are active or not, and it will obtain the payoff $\Pi_i$ as dictated by all the strategy pairs. If a player $i$ is inactive, it will not play any games with with its neighbors, and accordingly, its payoff $\Pi_i$ will be null. Furthermore, only active players are able to pass their strategy in agreement with the replicator equation, while inactive player are never able to replicate because their payoff is always zero (and hence never larger than the payoff of the neighbor).

The introduction of different activity patterns has two important effects on the evolutionary dynamics. First, the activation probabilities introduce a degree of heterogeneity to the evolutionary time scales, which in repeated settings can strongly affect the outcome since some players will be more active than others. Second, the additional layer of stochasticity (participation is probabilistic) in the spatiotemporal dynamics affects the correlation scales and can induce cascading effects that disrupt the formation of cooperative clusters. Based on these facts, it is clear that the distribution of activation probabilities in the population plays a crucial role. Accordingly, we consider power-law distributions of the form $Q(a) \sim a^{\tau}$, where $\tau$ is a free parameter that determines the slope of the line on a double-logarithmic scale. Furthermore, we consider separately (i) that $a_i$ are completely independent of $k_i$, and (ii) that $a_i = k_i/k_{max}$ where $k_{max}$ is the largest degree in the network. In the latter case, which applies specifically for SF networks, we thus assume a so-called topology-correlated activity pattern, while in the former case the activity of each player is uncorrelated with its degree. If topology-correlated activity applies, this means that hubs of the SF network will practically always be active, while low-degree players will participate only once per several time steps. Conversely, if activity and degree are uncorrelated, hubs too can experience long inactivation periods. As we will show next, this distinction has particularly important consequences for the overall evolutionary success of cooperative behavior.

\begin{figure}
\centering
\includegraphics[width=7.5cm]{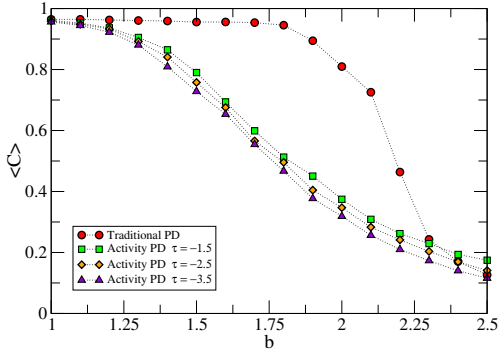}
\caption{Degree-uncorrelated activity patterns on SF networks impair the evolution of cooperation. Depicted is the average fraction of cooperators $\langle C \rangle$ in the population in dependence on the temptation to defect $b$, as obtained with the traditional prisoner's dilemma game (Traditional PD) and with three different $\tau$ values characterizing the activity-driven prisoner's dilemma game (Activity PD).}
\label{sfunc}
\end{figure}

We begin by presenting results obtained on SF networks when the activity pattern is uncorrelated with the degree of players. As shown in Fig.~\ref{sfunc}, for low and mild temptation the traditional version of the prisoner's dilemma game consistently returns significantly higher average levels of cooperation than activity-driven prisoner's dilemma games while, as $b$ increases,  differences tend to disappear. Moreover, it can be observed that the higher the heterogeneity in the distribution of activation probabilities (the more negative the value of $\tau$), the more impaired the evolution of cooperation. Although the differences that are generated by different values of $\tau$ are fairly small, they are nevertheless telling in terms of the negative effect inactivity has on cooperative behavior. In particular, if the activation probability is uncorrelated with the degree of players, chances are high that even some of the hubs will be inactive over extended periods of time. This has dire consequences for the evolution of cooperation since dormant cooperative hubs are unable to sustain a compact mass following of low-degree cooperators, which is the key mechanism behind elevated levels of cooperation \cite{santos_prl05}.

\begin{figure}
\centering
\includegraphics[width=7.5cm]{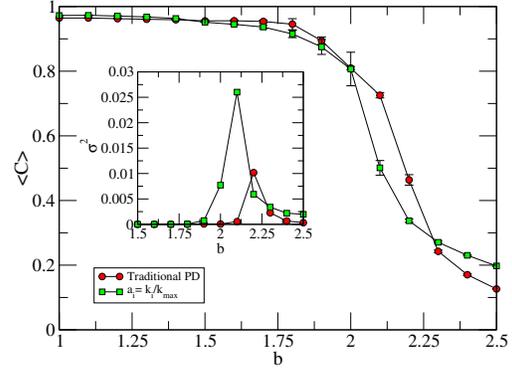}
\caption{Degree-correlated activity patterns on SF network do not impair the evolution of cooperation. The main panel shows the average fraction of cooperators $\langle C \rangle$ in the population in dependence on the temptation to defect $b$, as obtained with the traditional prisoner's dilemma game (Traditional PD) and with the activity-driven prisoner's dilemma game where $a_i=k_i/k_{max}$. The inset shows the variance in $\langle C \rangle$ ($\sigma^2$) as a function of $b$. Note that despite the similarities in $\langle C \rangle$, the difference in $\sigma^2$ is significant.}
\label{sfcor}
\end{figure}

However, if the activity pattern is correlated with the degree of players, such that hubs are assigned the highest $a_i$ values, the average level of cooperation in the stationary state becomes comparable to the outcome obtained with the traditional prisoner's dilemma game. Results presented in the main panel of Fig.~\ref{sfcor} attest clearly to this fact, and this in turn validates our explanation behind the impaired evolution of cooperation presented in Fig.~\ref{sfunc}. Indeed, if hubs are relatively unaffected by inactivity, the reasonable expectation is that the evolution of cooperation should not suffer. Yet the notable difference in variance in $\langle C \rangle$ as a function of $b$ shown in the inset of Fig.~\ref{sfcor} indicates that this is only partly the reason why the evolution of cooperation remains intact. Identical mean values but different variance indicate that the evolutionary dynamics leading to the stationary state is different, which we therefore explore further in more detail.

\begin{figure}
\centering
\includegraphics[width=7.5cm]{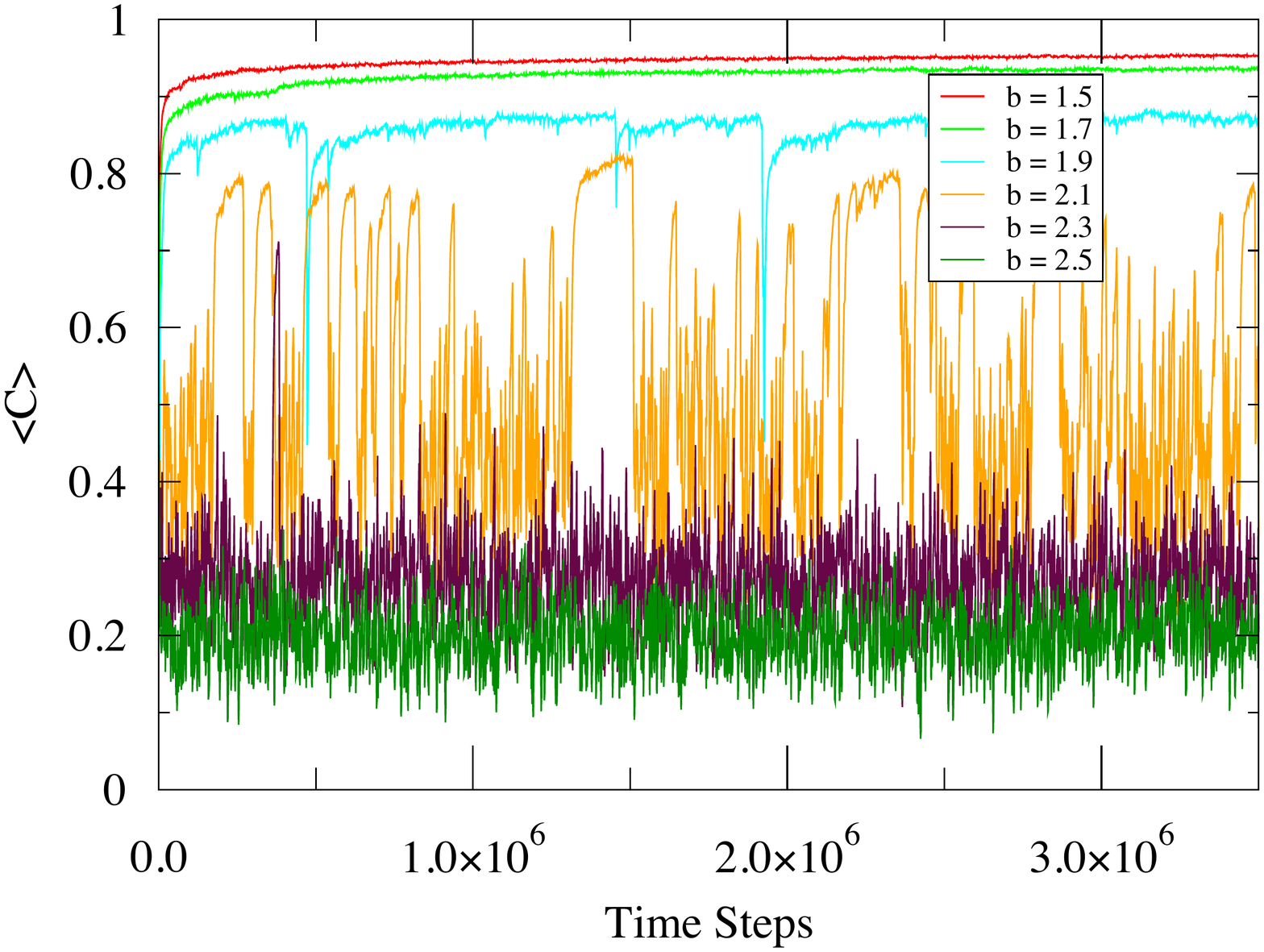}
\includegraphics[width=7.5cm]{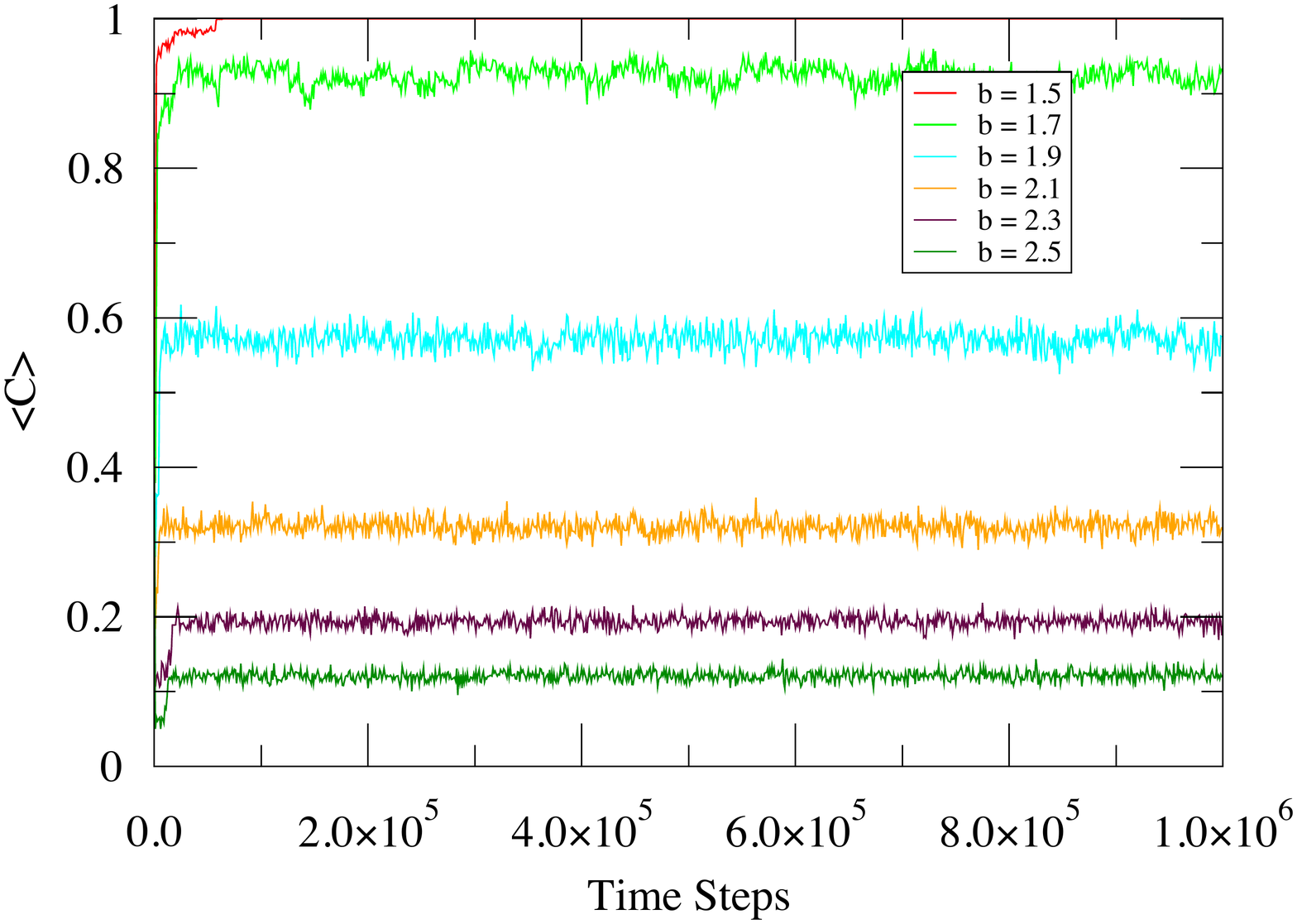}
\caption{Degree-correlated activity patterns on SF networks introduce cascading failures of cooperation. The drops in the fraction of cooperators are sudden, while the spatiotemporal reorganization of compact cooperative clusters, and thus the recovery, takes time (top panel). Comparatively, in the traditional version of the prisoner's dilemma game (bottom panel), the time evolution is smooth and without significant deviation over time. In both panels we show the time evolution of the fraction of cooperators $\langle C \rangle$, as obtained for different values of temptation to defect $b$ (see legend).}
\label{time}
\end{figure}

Results presented in Fig.~\ref{time} reveal the reason behind the difference in the variance reported in the inset of Fig.~\ref{sfcor}. The time courses of the fraction of cooperators in the presence of a heterogenous yet degree-correlated activity pattern are significantly more undulating than the time courses for the traditional prisoner's dilemma game. In comparison, the differences between the top and the bottom panel are most striking precisely in the intermediate region of $b$ values, where also the largest differences in $\sigma^2$ are recorded in the inset of Fig.~\ref{sfcor}. Elaborating on these differences, we find that the activity-driven game, for example at $b=1.9$, is subject to sudden drops in $\langle C \rangle$, which reoccur intermittently across the whole history of the game. The recovery periods are comparatively slow. Qualitatively similar instabilities have been observed before in the realm of evolutionary dynamics, for example in the prisoner's dilemma game with asymmetric influence \cite{kim_bj_pre02}, or in presence of imitation and mutation \cite{cavaliere_jtb12, yang_z_pone12}. In our case, however, the instabilities are due solely to the introduction of activation probabilities to individual players. Moreover with regards to the differences, we find that in-between the cascading failures, the level of cooperation is actually higher than in the traditional version of the game. Over time, the formerly described negative effect due to cascading failures and the latter positive effect cancel each other out, ultimately giving rise to the same average fraction of cooperators in both cases, as shown in the main panel of Fig.~\ref{sfcor}.

\begin{figure}
\centering
\includegraphics[width=7.5cm]{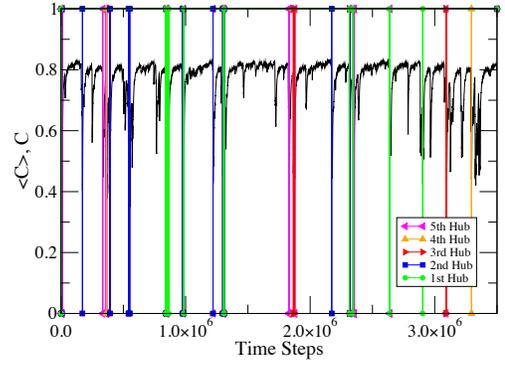}
\caption{Degree-correlated activity patterns on SF networks intermittently render cooperative hubs inactive, following which system-wide cascading failures of cooperation occur. Note that the lower the rank of the hub, the lower its activation probability, and thus the higher the probability that a $C$ state will be switched to a $D$ state. The drops have a characteristically asymmetric shape since defectors can invade immediately without prior spatiotemporal organization, while cooperators require more time to reorganize into compact cluster and regain lost ground. Depicted is the time evolution of $\langle C \rangle$ and state $C$ ($C=1$ cooperation, $C=0$ defection) of the five most connected nodes in the network (see legend), as obtained for $b= 2.0$. Here $a_i=k_i/k_{max}$.}
\label{state}
\end{figure}

\begin{figure}
\centering
\includegraphics[width=7.5cm]{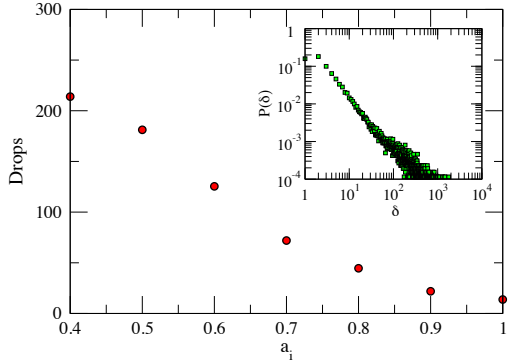}
\caption{The switch from cooperation is the more common the lower the activation probability. Depicted is the number of $C \to D$ changes for the $100$ most connected players in dependence on their activation probability $a_i$. The inset shows the distribution of times $\delta$ between consecutive drops in $\langle C \rangle$. Here a drop is defined as a monotonous decrease in $\langle C \rangle$ by at least $0.005$ (50 players on a network of size $N=10^4$). These statistics have been collected after a transient period of $10^5$ time steps when the entire simulation run for $3.5 \times 10^6$ time steps in the degree-correlated activity model with $b=2.0$. Points in the main panel are the average over intervals of $0.1$ in $a_i$.}
\label{stats}
\end{figure}

To obtain further insights into the intriguing temporal complexity of the evolution of cooperation driven by heterogeneous activity patterns, we show in Fig.~\ref{state} how the state of the main hubs of the SF network varies over time, and how these changes correlate with the drops in $\langle C \rangle$. In essence, the inactivity increases the chance of a cooperative hub to become invaded by a defector. The lower the rank of the hub, the higher this probability. We quantify this explicitly in the main panel of Fig.~\ref{stats}, where the drop is the more likely the lower the activation probability $a_i$. As soon as a hub changes from state $C$ to state $D$, the drop in $\langle C \rangle$ is unavoidable and sudden because defectors invade cooperators fast and do not require any spatiotemporal organization to do so. When the hub is reoccupied by a cooperator, however, the recovery takes a while longer because cooperators, unlike defectors, do require the formation of compact clusters to spread (win back lost ground in this case). To confirm that these sudden drops in the level of cooperation are indeed due to the activation patterns, we show in the inset of Fig.~\ref{stats} the distribution of times between consecutive drops. Evidently, the power-law distribution is a direct consequence of the imposed activity pattern. But as long as cooperators do occupy the hubs, in-between the cascading failures, the lower-degree cooperators benefit from the inactivity since they suffer from it much less than defectors. More precisely, a low-degree cooperator will be unable to spread regardless of its activity, but a low-degree defector is likely to spread, if only it is active enough. As a result, the phase between the drops is characterized by higher $\langle C \rangle$ levels, which together with the intermittent drops results in more or less the same final average level of cooperation as recorded in the traditional prisoner's dilemma game.

Lastly, we show the results obtained on ER random networks, where like in the case of degree-correlated activity patterns on SF networks shown in Fig.~\ref{sfcor}, different activity patterns also do not impair the evolution of cooperation. This confirms that the particularities of the activity pattern play a particularly important role on strongly heterogeneous interaction networks, while the evolution on lattices, small-world and random networks is comparatively little affected. Since notable hubs on such networks are absent, the activity pattern, even if uncorrelated with degree (the distinction between degree-uncorrelated and degree-correlated activity actually plays no role), does not induce such sudden and fast drops in the fraction of cooperators as we have reported in Figs.~\ref{time} and \ref{state}. The clustering of cooperators is of course occasionally disrupted by inactive players that quickly fall prey to defectors, but this just counteracts what the defectors lose by having an overall lower potency to invade due to the imposed periods of inactivity. On average, the cooperation level thus stays the same, or perhaps slightly tilted in favor of cooperators at intermediates values of $b$.

\begin{figure}
\centering
\includegraphics[width=7.5cm]{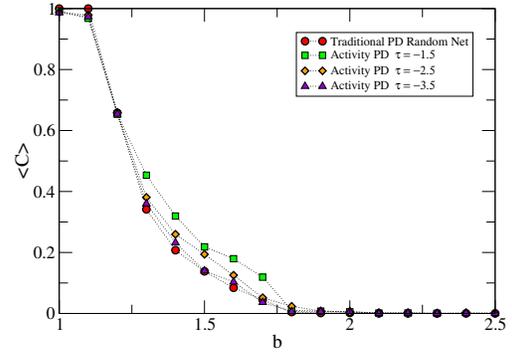}
\caption{Different activity patterns on ER random networks do not impair the evolution of cooperation. Depicted is the average fraction of cooperators $\langle C \rangle$ in the population in dependence on the temptation to defect $b$, as obtained with the traditional prisoner's dilemma game (Traditional PD) and with three different $\tau$ values characterizing the activity-driven prisoner's dilemma game (Activity PD).}
\label{random}
\end{figure}

Summarizing, we have studied the impact of diverse activity patterns on the evolution of cooperation in evolutionary social dilemmas. We have focused on the weak prisoner's dilemma on random and scale-free networks. The different activity patterns have been implemented through individual activation probabilities, according to which players were either active or inactive during a particular round of the game. While active agents played the game with all their neighbors and obtained their payoffs accordingly, inactive players did not play and thus ended up with a null payoff. Consequently, only active players were able to pass their strategies to their neighbors, while inactive players were unable to replicate.

We have shown that degree-uncorrelated activity patterns on SF networks significantly impair the evolution of cooperation. This is due to the fact that elevated levels of cooperation on SF networks require fully functional, always active hubs. Thus, in case one or several hubs were assigned a low activation probability, the average level of cooperation in the population dropped significantly. As is well known, cooperative hubs act as important centers of sizable and compact communities of cooperators \cite{santos_prl05}. And if these hubs are rendered inactive for extended periods of time, then obviously their support for extensive cooperative communities to emerge is no longer provided. Conversely, when we have considered degree-correlated activity patterns on SF networks, such that predominantly low-degree nodes were affected by extensive periods of inactivity, we have found that the evolution of cooperation is, at least on average, not negatively affected. Although sudden and steep drops in the level of cooperation still occurred, they were compensated by an overall higher level of cooperation in-between the cascading failures. Interestingly, we have demonstrated that, as long a cooperators do occupy the hubs, the
lower-degree cooperators benefit from the inactivity since they suffer from it much less than defectors. In particular, defectors are unable to utilize on their inherently higher payoffs, which lastly balances out the asymmetric dynamic instabilities that emerge due to intermittent blackouts of cooperative hubs.

We have further corroborated this conclusions by studying the impact of different activity patterns on random ER networks. In this case, cooperators rely predominantly on traditional network reciprocity \cite{nowak_n92b}, rather than on the enhanced version of the same mechanism that works on SF networks due to the strong heterogeneity of the degree distribution. Accordingly, the distinction between degree-uncorrelated and degree-correlated activity patterns plays no role. Indeed, we have shown that even a random assignment of activation probabilities fails to lower the average level of cooperation in the population. As much as cooperators lose by having their clustering process disrupted by occasionally inactive players that quickly fall prey to defectors, the defectors lose by having an overall lower potency to invade due to the imposed periods of inactivity. On average, and regardless of correlations between inactivity and the degree of players, the cooperation level thus stays the same.

We conclude by noting that our study is only the first step towards the introduction of different activity patterns in evolutionary games, which ought to properly take into account the diversity of not only the applied strategies, but also the diversity of individual activity levels of players \cite{capraro_pone13, capraro_sr14}. However, already the simplest consideration has revealed remarkable complexity in the evolutionary dynamics, involving cascading failures that manifest as asymmetric dynamic instabilities in the time course of the evolution of the two competing strategies. An interesting direction for future research certainly involves human experiments with an exit option, as studied before in dictator games \cite{dana_07, capraro_a14}. Future simulations might involve players being able to change their activity patterns dynamically over time, for example depending on their evolutionary success in previous rounds of the game. Relative times scales in evolutionary dynamics \cite{roca_prl06} could also play an important role, in that the typical time for an ``activation change'' might be different from the strategy change. To explore the consequences of these options appears to be an exciting venture with many relevant implications, and we hope that this letter will motivate further research along this line in the near future.

\acknowledgments  This work has been partially supported by the National Natural Science Foundation of China (NSFC) through Grant No. 61374169 to C-YX; MINECO through Grant FIS2011-25167 to YM; Comunidad de Arag\'on (Spain) through a grant to the group FENOL to YM;  the EC Proactive project MULTIPLEX (contract no. 317532) to SM, YM. SM is supported by the MINECO through the Juan de la Cierva Program. MP acknowledges financial support from the Deanship of Scientific Research, King Abdulaziz University (Grant 76-130-35-HiCi).

\end{document}